\documentclass[final,5p,twocolumn,times,11pt]{elsarticle}
\usepackage{graphicx,color,amsmath,amssymb}
\usepackage{bm}
\usepackage[colorlinks=true, pdfstartview=FitV, linkcolor=red, citecolor=blue, urlcolor=blue]{hyperref}

\usepackage{epstopdf}
\advance\voffset -0.2in
\biboptions{sort&compress}

\begin{document}

\begin{frontmatter}
\title{Rapidity dependence of elliptic and triangular flow in proton-nucleus collisions\\from collective dynamics 
}

\author[agh]{Piotr Bo\.zek}
\ead{bozek@fis.agh.edu.pl}

\author[agh]{Adam Bzdak}
\ead{bzdak@fis.agh.edu.pl}

\author[gm]{Guo-Liang Ma}
\ead{glma@sinap.ac.cn}

\address[agh]{AGH University of Science and Technology, Faculty of Physics and Applied Computer Science, PL-30-059 Krak\'ow, Poland}
\address[gm]{Shanghai Institute of Applied Physics, Chinese Academy of Sciences, Shanghai 201800, China}

\begin{abstract}
The rapidity dependence of elliptic, $v_2$, and triangular, $v_3$, flow coefficients in proton-nucleus (p+A) collisions is predicted in hydrodynamics and in a multi-phase transport model (AMPT). We find that $v_n$ ($n=2,3$) on a nucleus side is significantly larger than on a proton side and the ratio between the two, $v_n^{\rm Pb}/v_n^{\rm p}$, weekly depends on the transverse momentum of produced particles.  
\end{abstract}
\end{frontmatter}


\section{Introduction}
\label{sec: introduction}

Significant second and third harmonics have been 
observed in the long-range 
azimuthal correlations of particles emitted in
ultra-relativistic p+Pb collisions at the LHC \cite{CMS:2012qk,Abelev:2012ola,Aad:2013fja}
and d+Au collisions at RHIC \cite{Adare:2013piz}. The results can be interpreted as due to collective flow of particles in the framework of hydrodynamic models \cite{Bozek:2011if,Bozek:2013uha,Bzdak:2013zma,Qin:2013bha,Kozlov:2014fqa,Werner:2013ipa,Nagle:2013lja} or in the cascade AMPT model\footnote{As shown in Ref. \cite{He:2015hfa}, the AMPT model generates the signal mostly due to the escape mechanism which presumably differs from hydrodynamics.} \cite{Ma:2014pva,Bzdak:2014dia,Koop:2015wea}. A different approach connects the observed particle correlations with saturation 
effects in the initial state of the collision
\cite{Dumitru:2014dra,Dumitru:2014yza,Kovchegov:2012nd,Dusling:2013oia}. 
Observables related to rapidity dependence of the bulk quantities can be used to disentangle between the two mechanisms. An example of such observable is the average transverse momentum of produced particles as a function of (pseudo)rapidity. The average transverse momentum is predicted to be larger 
on the Pb-going than on the p-going side in the hydrodynamic model \cite{Bozek:2013sda}, while the reverse is expected in the color glass condensate \cite{Gelis:2010nm} (CGC) approach. The transverse size of the fireball is larger (and it lives longer) on the
Pb-side which results in a stronger collective flow during the evolution, leading to not only a larger transverse flow (and larger $\langle p_\perp  \rangle$), but also a stronger 
elliptic and triangular flow. Such an asymmetry of integrated elliptic flow in p- and Pb-going sides has been observed by the CMS collaboration \cite{GranierdeCassagnac:2014jha}. 
In this letter 
we present a calculation of the 
relative strength of the elliptic and the triangular flow in the Pb-going and p-going sides 
as a function of the transverse momentum  
 in the 3+1 dimensional (3+1D)
 viscous hydrodynamic model and in the AMPT model.  In both cases we
 observe a significant increase of the elliptic and the triangular flow coefficients for
 rapidities corresponding to the Pb-nucleus direction. Our predictions extend to larger rapidities than measured by the CMS collaboration, and correspond to acceptance of the ALICE muon spectrometer.

In the next section we present our main results. In section 3 we offer some comments and we conclude the paper in section 4.

\section{Results}
\label{sec:results}

In this Section we present the elliptic and triangular flow coefficients in the proton- and the nucleus-going directions calculated in the 3+1D hydrodynamics and the AMPT model.

\subsection{Hydrodynamics}

The initial density for the hydrodynamic evolution 
is calculated in the Glauber Monte Carlo model. The entropy is deposited at the 
nucleon-nucleon collision points with a Gaussian profile in the transverse plane
\cite{Bozek:2013uha}.   
The 3+1D hydrodynamic calculations are performed event-by-event, with shear viscosity $\eta/s=0.08$ and bulk viscosity $\eta/s=0.04$ 
for $T<170$ MeV. At the freeze-out temperature of  $150$ MeV, particles are emitted
statistically \cite{Chojnacki:2011hb}, including corrections due to bulk
 and shear viscosity in the Cooper-Frye formula \cite{Bozek:2009dw}.
The
 hydrodynamic simulations reproduce fairly well
 the measured elliptic 
and triangular flow
\cite{Bozek:2013uha}, the mass hierarchy  of the
 elliptic flow coefficient and of the  average transverse momentum 
of identified particles \cite{Bozek:2013ska}, and the
 interferometry radii \cite{Adam:2015pya}.

The centrality $0$-$20$\% is defined as events with the number 
of wounded nucleons $N_w \ge 13$.
Charged particles are analysed in three bins, the forward (Pb-going side) 
$2.5<\eta<4$ 
and backward (p-going side) $-4<\eta<-2.5$, and the central bin $|\eta|<1$.
The central bin defines the reference event-plane for charged particles with 
$0.25<p_\perp<5$ GeV. The flow coefficients $v_n\{2\}(p_\perp)$ 
for charged particles in the 
forward and backward bins are calculated with respect
 to the reference particles  from the central bin.
 
\begin{figure}
\includegraphics[scale=0.45]{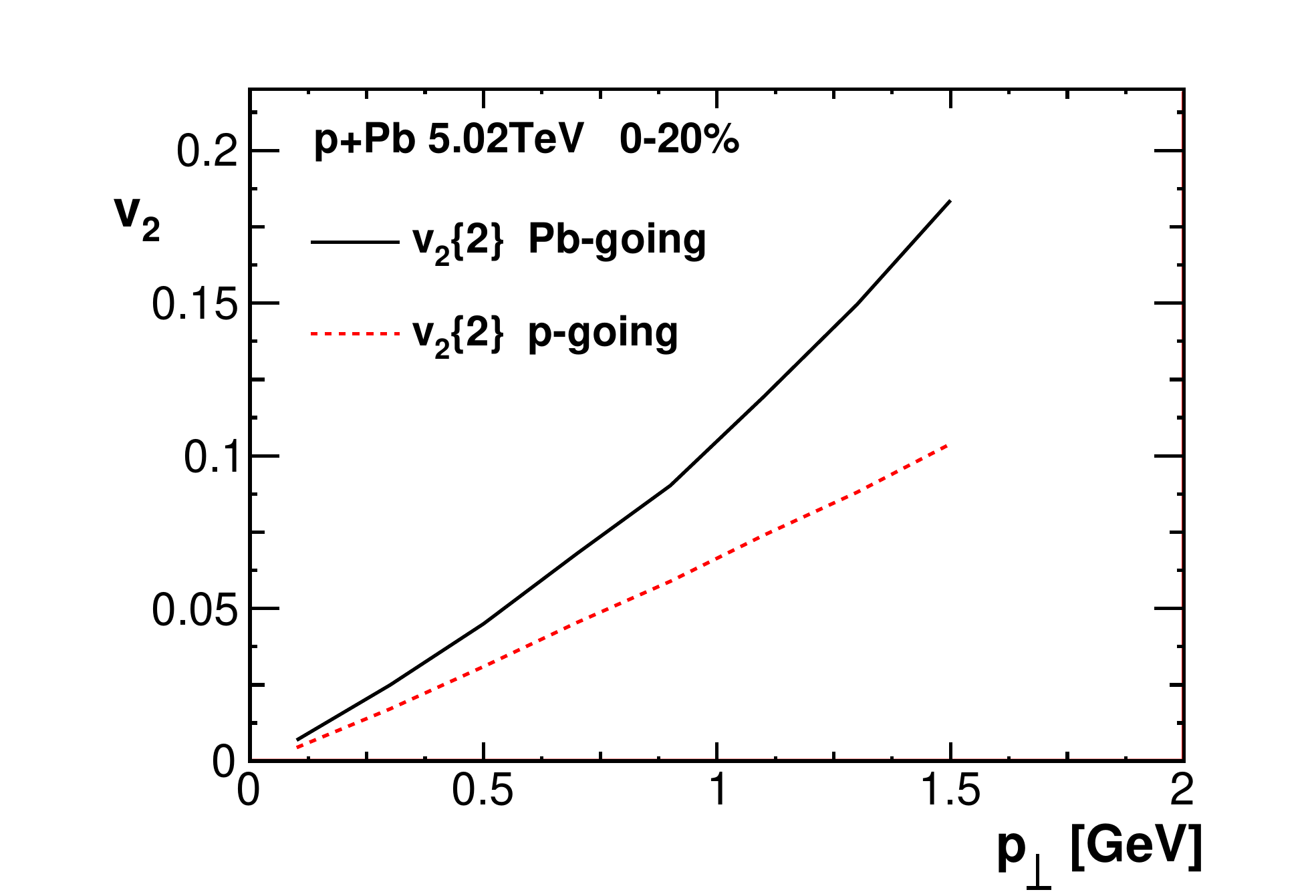}
\caption{The elliptic flow coefficient on the proton- ($-4<\eta<-2.5$) (dashed line) and  the nucleus-going ($2.5<\eta<4$) (solid line) sides in $0-20\%$ p+Pb collisions at $\sqrt{s}=5.02$ TeV, as a function of the transverse momentum, $p_\perp$, from 3+1D hydrodynamics.}
\label{fig:hydrov2}
\end{figure}
\begin{figure}
\includegraphics[scale=0.45]{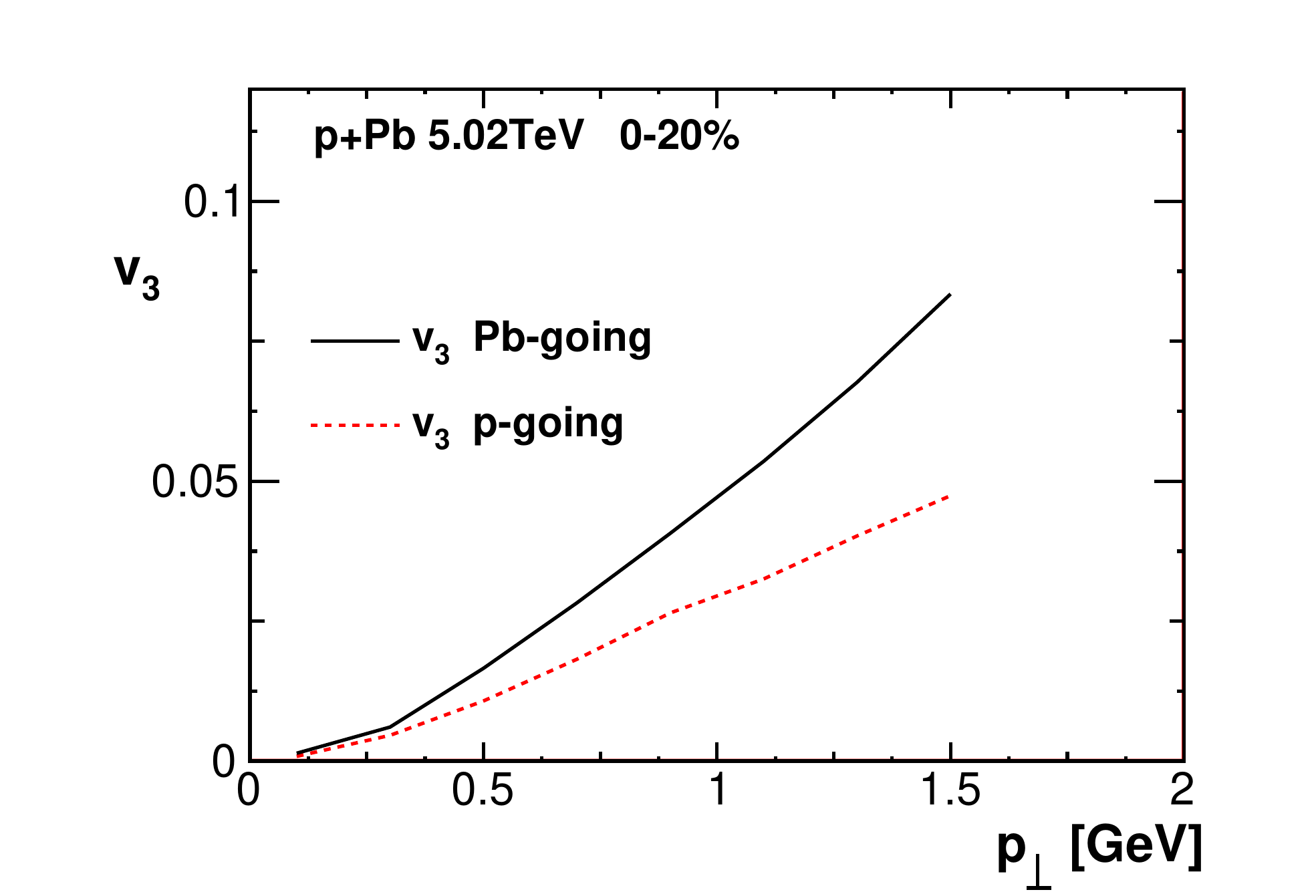}
\caption{Same as figure \ref{fig:hydrov2} but for the triangular flow.}
\label{fig:hydrov3}
\end{figure}
\begin{figure}
\includegraphics[scale=0.45]{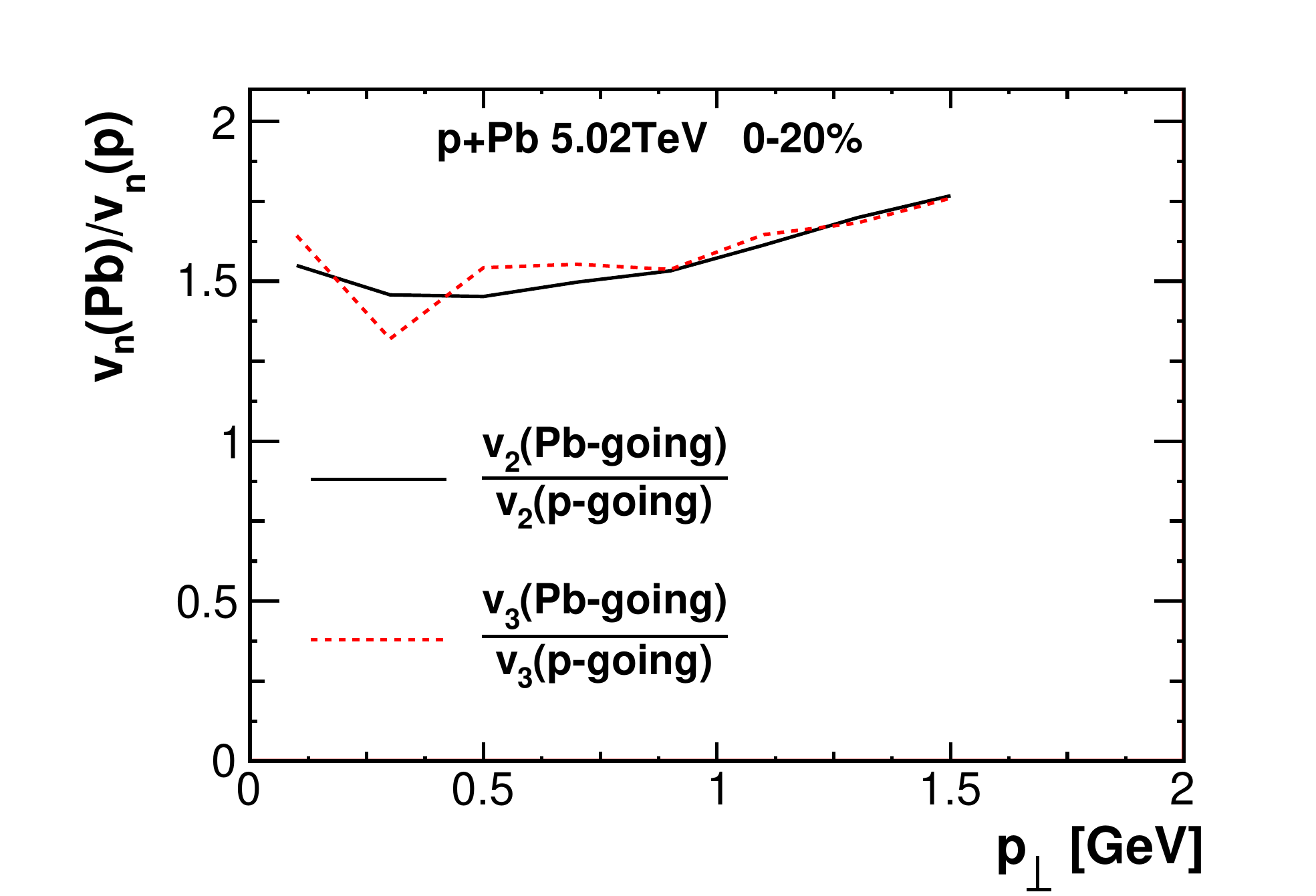}
\par
\vspace{-5mm}
\caption{The ratio of the Pb-going to p-going values 
of the elliptic (solid line) and triangular (dashed line) flow coefficients, from the 3+1D hydrodynamic calculation.}
\label{fig:hydroratio}
\end{figure}

The elliptic and triangular flow  as a function of 
the transverse momentum is larger on the nucleus-going side 
(Figs. \ref{fig:hydrov2} and \ref{fig:hydrov3}). Since the average transverse momentum is expected to be larger on the 
nucleus-going side \cite{Bozek:2013sda}, the difference for 
 integrated flow coefficients is predicted to be even larger. The origin of the effect
in the hydrodynamic model  can be linked to the 
longer lifetime of the fireball on the nucleus going side,
 which results in a stronger built up of the collective
 flow. Moreover, for rapidities where the freeze-out happens earlier we expect much stronger viscosity correction at freeze-out, reducing the flow coefficients \cite{Teaney:2003kp}. As shown in Fig. \ref{fig:hydroratio}, the ratio of the flow coefficients calculated 
in the forward and backward rapidity bins weakly depends on $p_\perp$, in the range where the hydrodynamic model applies.

\subsection{A multi-phase transport model}

The AMPT model with the string melting mechanism proved to be very effective in describing various features of p+Pb, d+Au and the high-multiplicity p+p interactions data \cite{Ma:2014pva,Bzdak:2014dia,Koop:2015wea}.\footnote{We note that approximately $1-2$ elastic collisions per partons suffice to describe the p+Pb data \cite{Ma:2014pva,Bzdak:2014dia}.} The model is initialized with soft strings (soft particles) and minijets (hard particles) from HIJING \cite{Wang:1991hta}. In the string melting scenario both strings and minijets are converted into quarks and anti-quarks that subsequently undergo elastic scatterings with a given cross-section, $\sigma$, which is a free parameter.\footnote{The AMPT model with $\sigma=0$ is equivalent to HIJING (plus hadronic transport \cite{Li:1995pra}).} It was found that a cross-section of $1.5-3$ mb is sufficient to reproduce the data in p+p and p+Pb collisions at the LHC \cite{Ma:2014pva,Bzdak:2014dia}, and d+Au interactions at RHIC \cite{Koop:2015wea}. In this paper we choose $\sigma=3$ mb.    

\begin{figure*}
\begin{center}
\includegraphics[scale=0.7]{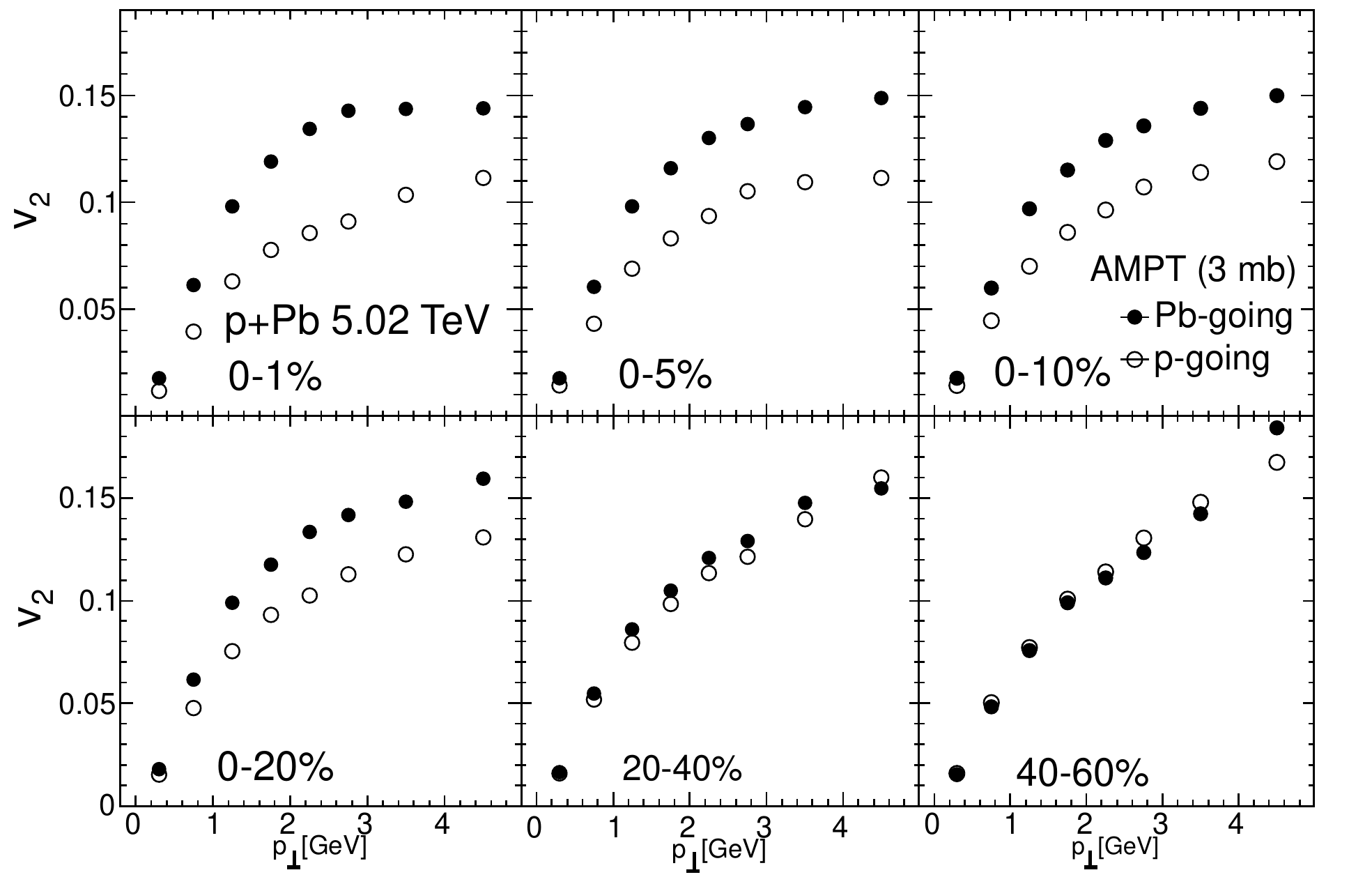}
\end{center}
\par
\vspace{-5mm}
\caption{The AMPT model (with string melting) results for the elliptic flow coefficients on a proton ($-4<\eta<-2.5$) and a nucleus ($2.5<\eta<4$) sides, $v_2^{\rm p}$ and $v_2^{\rm Pb}$,  for various centrality classes in p+Pb interactions at $\sqrt{s}=5.02$ TeV, as a function of the transverse momentum, $p_\perp$. In this plot jets contribute to $v_2$ for higher values of $p_\perp$.}
\label{fig:ampt_v2}
\end{figure*}

In Fig. \ref{fig:ampt_v2} we present $v_2$ in the proton- ($-4<\eta<-2.5$) and the nucleus-going ($2.5<\eta<4$) directions as a function of the transverse momentum, $p_\perp$.
We performed our calculations in two different ways. In the first method we calculated three two-particle correlation functions between two bins (i) $-4<\eta<-2.5$ and $|\eta|<1$, (ii) $2.5<\eta<4$ and $|\eta|<1$, and (iii) $-4<\eta<-2.5$ and $2.5<\eta<4$.\footnote{In all cases we have large enough rapidity separation between bins allowing to neglect correlations from jet cones, etc.} In this way we can extract $v_2^{\rm p}v_2^{\rm mid}$, $v_2^{\rm Pb}v_2^{\rm mid}$ and $v_2^{\rm p}v_2^{\rm Pb}$ what allows to calculate $v_2^{\rm p}$ and $v_2^{\rm Pb}$ separately.   

In the second method we extract $v_2^{\rm p}v_2^{\rm mid}$ and $v_2^{\rm Pb}v_2^{\rm mid}$ as above however, in this case we extract $v_2^{\rm mid}$ calculating $v_2$ in $|\eta|<1.2$ with the rapidity gap between particles being two units of rapidity. This allows to extract $v_2^{\rm mid}$ with a good approximation and in fact we checked that both methods lead to practically indistinguishable results. In the following we show the results obtained using the latter method.

As seen in Fig. \ref{fig:ampt_v2}, $v_2^{\rm Pb}$ is larger that $v_2^{\rm p}$ (except peripheral collisions) for all calculated values of $p_\perp$. In peripheral collisions the single particle $\eta$ distribution, $dN/d\eta$ becomes almost symmetric in $\eta$ leading to $v_2^{\rm p} \approx v_2^{\rm Pb}$. 

\begin{figure*}
\begin{center}
\includegraphics[scale=0.7]{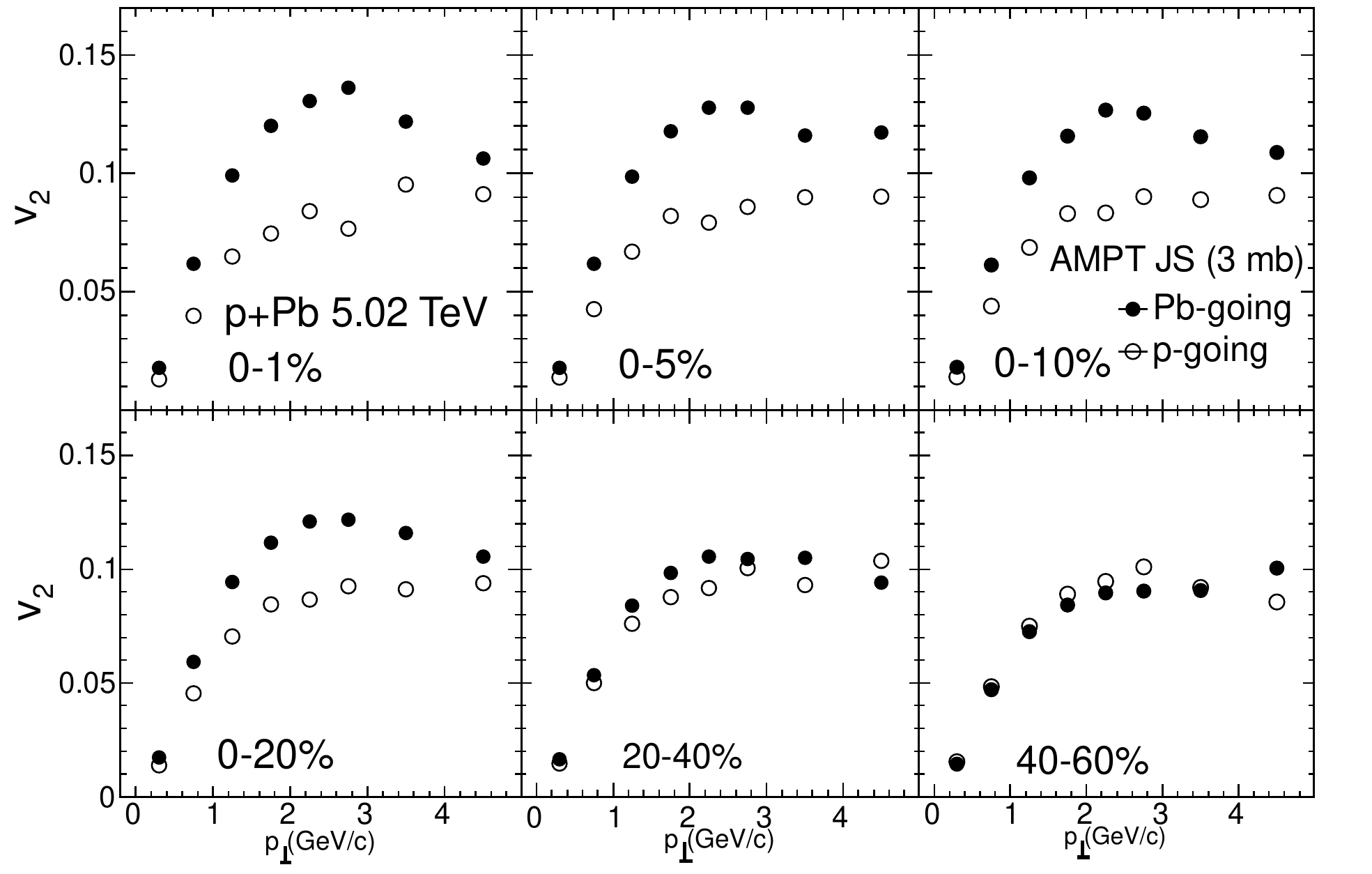}
\end{center}
\par
\vspace{-5mm}
\caption{Same as Fig. \ref{fig:ampt_v2} except jets are subtracted (JS) from the two-particle azimuthal correlation function reducing both $v_2^{\rm p}$ and $v_2^{\rm Pb}$ for higher values of $p_\perp$.}
\label{fig:ampt_js_v2}
\end{figure*}

It is important to see how jets influence both $v_2^{\rm p}$ and $v_2^{\rm Pb}$. We subtracted jets by randomizing the azimuthal angles between produced jets. In this case jets do not contribute to the two-particle correlation function at $\Delta\phi=\pi$ and consequently do not contribute to the extracted values of $v_2$ (provided we have large enough rapidity gap between bins). As seen in Fig. \ref{fig:ampt_js_v2} this procedure modifies $v_2$ at large transverse momenta however, all qualitative features remain unchanged.\footnote{We also checked our results using a different method of jet subtraction. We calculated the two-particles correlation functions for $\sigma=3$ and $\sigma=0$ mb. The latter has no contribution from collective physics. Finally we calculated $\sqrt{v_2(3 \rm mb)^2 - v_2(0 \rm mb)^2}$ that is sensitive to collective physics only. We found that both methods lead to practically the same results.} 

\begin{figure*}
\begin{center}
\includegraphics[scale=0.7]{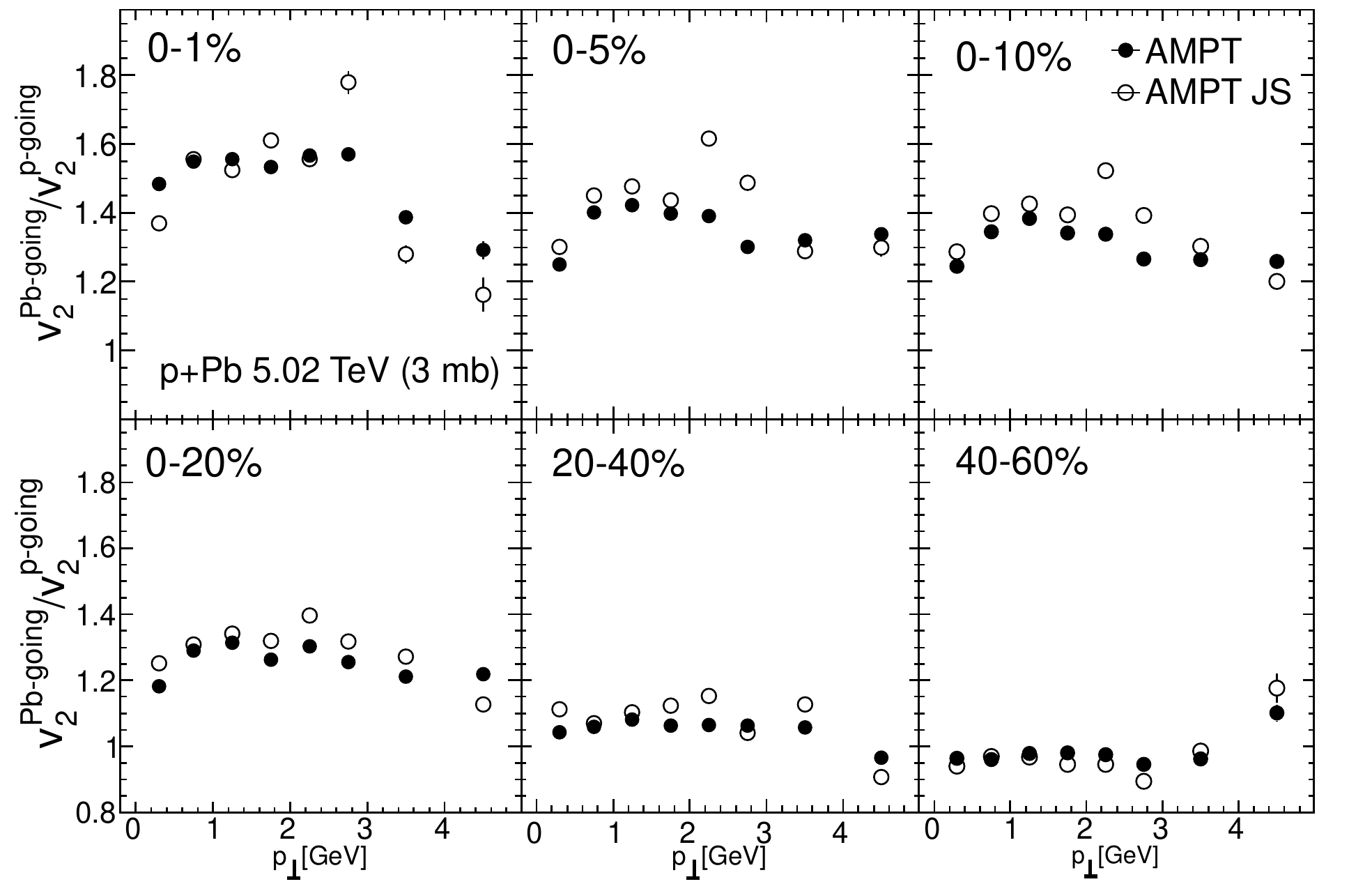}
\end{center}
\par
\vspace{-5mm}
\caption{The AMPT and AMPT JS (jets subtracted) results for the ratio $v_2^{\rm Pb}(p_\perp)/v_2^{\rm p}(p_\perp)$ as a function of the transverse momentum $p_\perp$.}
\label{fig:ampt_v2_ratio}
\end{figure*}

In Fig. \ref{fig:ampt_v2_ratio} we show the ratio between $v_2^{\rm Pb}$ and $v_2^{\rm p}$ as a function of transverse momentum calculated in the AMPT and AMPT JS (jets subtracted) models. 

\begin{figure}[t]
\includegraphics[scale=0.45]{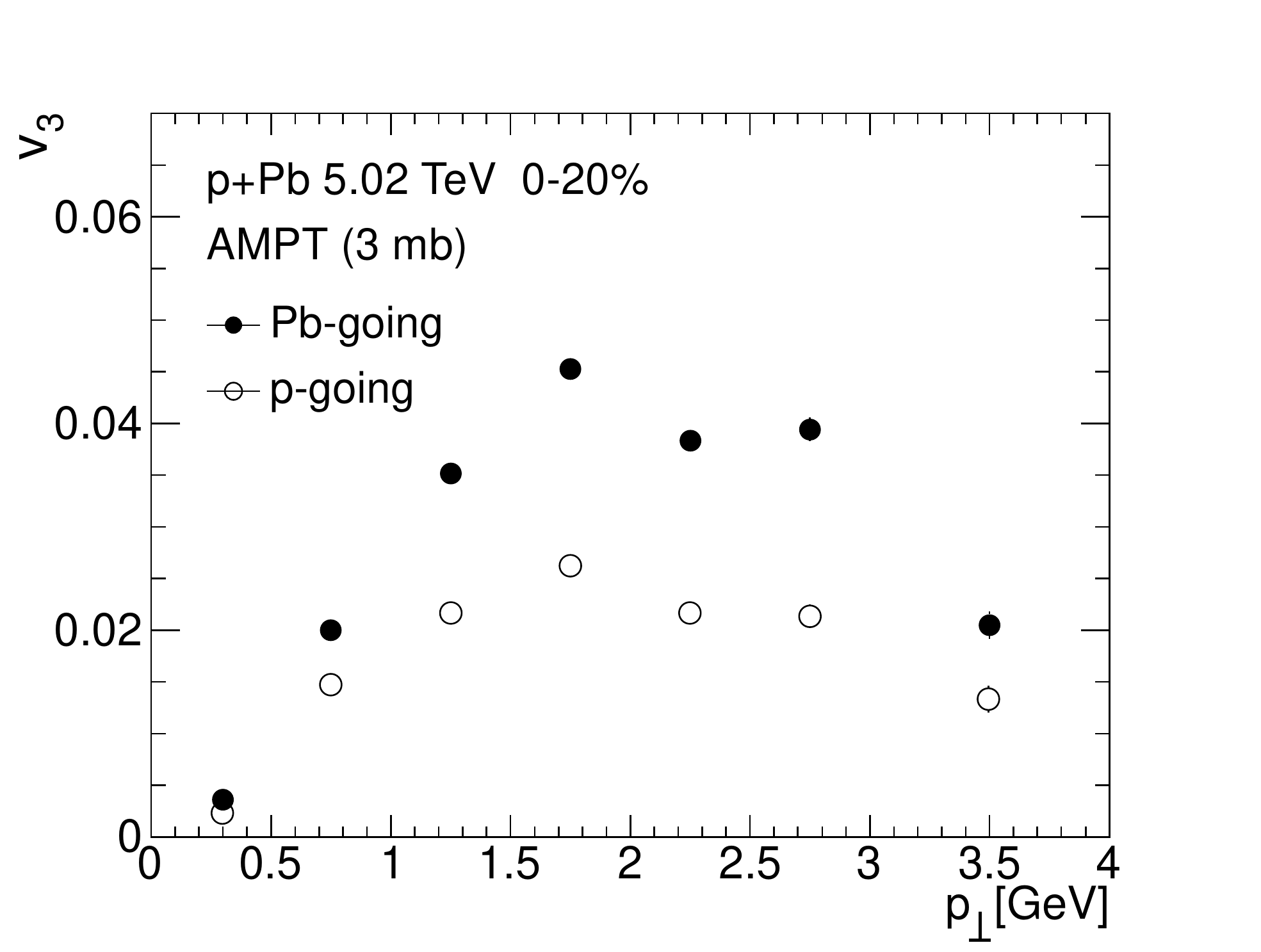}
\par
\vspace{-5mm}
\caption{Same as figure \ref{fig:hydrov3} but from the AMPT model.}
\label{fig:ampt_v3}
\end{figure}

\begin{figure}[t]
\includegraphics[scale=0.45]{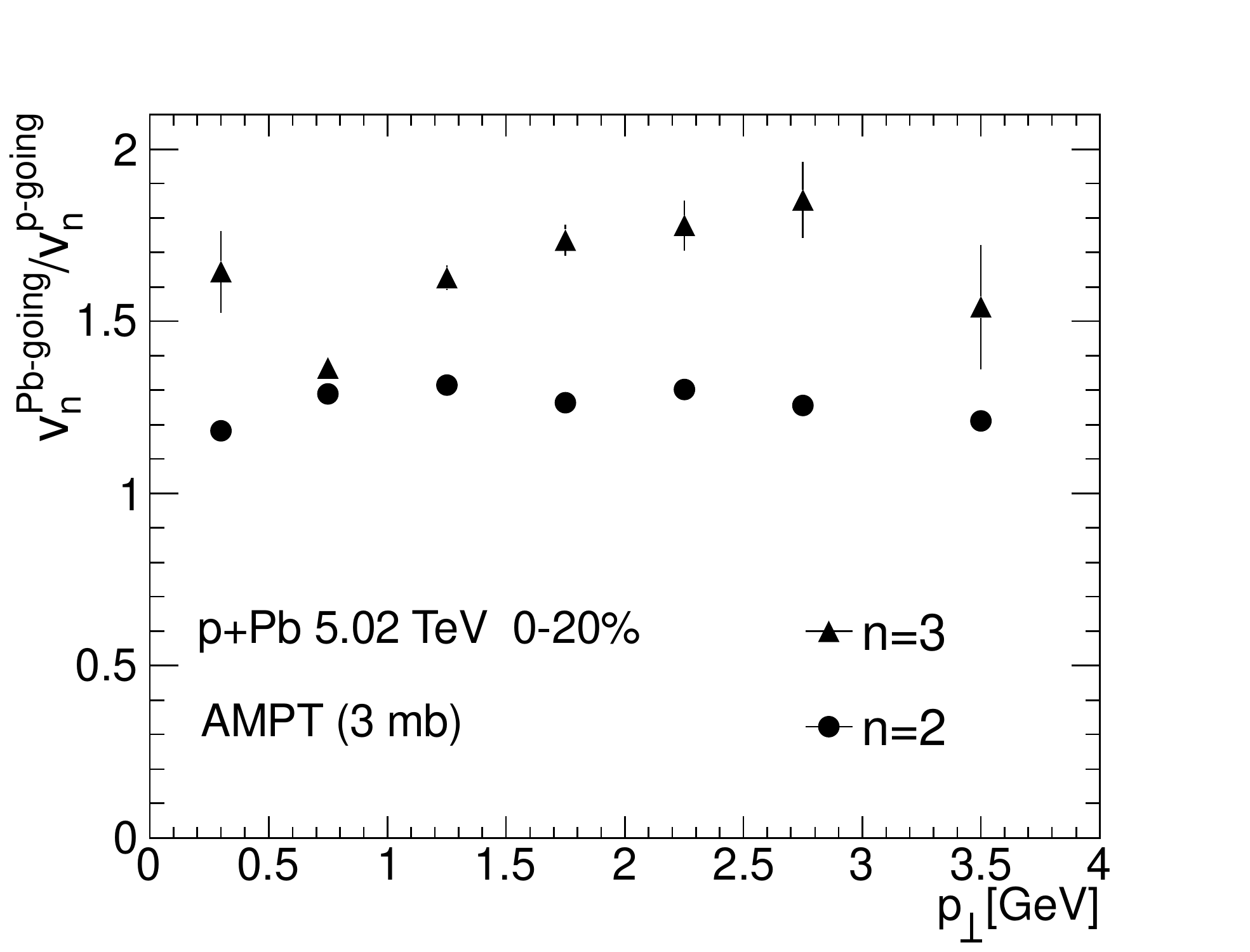}
\par
\vspace{-5mm}
\caption{Same as figure \ref{fig:hydroratio} but from the AMPT model.}
\label{fig:ampt_v3_ratio}
\end{figure}

Finally we performed our calculations for $v_3$. In Fig. \ref{fig:ampt_v3} we show $v_3^{\rm Pb}$ and $v_3^{\rm p}$ as a function of transverse momentum in $0-20\%$ p+Pb collisions at $\sqrt{s}=5.02$ TeV calculated in the AMPT model. In Fig. \ref{fig:ampt_v3_ratio} we present the ratios of $v_2^{\rm Pb}/v_2^{\rm p}$ and $v_3^{\rm Pb}/v_3^{\rm p}$ for $0-20\%$ centrality class in the AMPT model, where a larger ratio for $v_3$ than $v_2$ is seen.\footnote{We checked that as expected the AMPT JS model (jets subtracted) gives almost identical results.} 

\section{Comments}
\label{sec:comments}

Several comments are in order.

We repeated our calculations using different definitions of centrality classes. We checked several possibilities including: cuts of the multiplicity distributions in $|\eta|<1$, $2.8<\eta<5.1$ and $5<\eta<6$, and cuts in the number of wounded nucleons, $N_{\rm part}$. As expected, we found some quantitative differences (on the level of $20\%$ for $0-20\%$ centrality class) however, all qualitative features remained unchanged.

It would be interesting to perform analogous calculations in the color glass condensate framework. $v_2$ on a nucleus side is driven by large $x$ partons in a nucleus and small $x$ partons in a proton with the opposite situation for $v_2$ on a proton side. Consequently in CGC we expect a nontrivial dependence of $v_2$ on rapidity in p+Pb collisions and it is plausible that $v_2(\eta)$ could serve as the decisive test of the initial vs. the final state effects.

\section{Conclusions}
\label{sec:conclusions}

In conclusion, we predicted the rapidity dependence of elliptic and triangular flow coefficients in p+Pb collisions at the LHC energy using the AMPT and 3+1D hydrodynamics models. We found that both $v_2$ and $v_3$ in central collisions are significantly larger on a nucleus side ($2.5<\eta<4$) than on a proton side ($-4<\eta<-2.5$) and the ratio between the two, $v_n^{\rm Pb}(p_\perp)/v_n^{\rm p}(p_\perp)$, weakly depends on the transverse momentum of produced particles. The signal is somehow larger in hydrodynamics than in the AMPT model. We also predicted the centrality dependence of the effect and found that already for $40-60\%$ centrality class the ratio is consistent with unity.
It was further observed that the ratio weakly depends on various methods of centrality definition in p+Pb (for $0-20\%$ centrality class). Finally, we performed our calculations with and without jet contribution (by randomizing azimuthal angle between produced jets in AMPT) and found very little effect on the ratio whereas the individual $v_2       $ coefficients are obviously strongly modified at larger $p_\perp$. It would be interesting to perform analogous calculations in the initial state models of p+A interactions, where a nontrivial $v_2$ dependence on (pseudo)rapidity is expected. We hope our results will provide a stronger test of the collective dynamics in p+A collisions.   

\bigskip 

\section*{Acknowledgments}

Supported by the Ministry of Science and Higher Education (MNiSW), by PL-Grid Infrastructure, by founding from the Foundation for Polish Science, and by the National Science Centre, Grant No. DEC-2012/06/A/ST2/00390 and UMO-2013/09/B/ST2/00497. G.-L. M. is supported by the Major State Basic Research Development Program in China under Grant No. 2014CB845404, the National Natural Science Foundation of China under Grants No. 11175232, No. 11375251, and No. 11421505.

\bigskip

\bibliography{hydr}

\begin{thebibliography}{29}
\expandafter\ifx\csname natexlab\endcsname\relax\def\natexlab#1{#1}\fi
\providecommand{\url}[1]{\texttt{#1}}
\providecommand{\href}[2]{#2}
\providecommand{\path}[1]{#1}
\providecommand{\DOIprefix}{doi:}
\providecommand{\ArXivprefix}{arXiv:}
\providecommand{\URLprefix}{URL: }
\providecommand{\Pubmedprefix}{pmid:}
\providecommand{\doi}[1]{\href{http://dx.doi.org/#1}{\path{#1}}}
\providecommand{\Pubmed}[1]{\href{pmid:#1}{\path{#1}}}
\providecommand{\bibinfo}[2]{#2}
\ifx\xfnm\relax \def\xfnm[#1]{\unskip,\space#1}\fi
\bibitem[{Chatrchyan et~al.(2013)}]{CMS:2012qk}
\bibinfo{author}{S.~Chatrchyan}, et~al. (\bibinfo{collaboration}{CMS
  Collaboration}), \bibinfo{journal}{Phys. Lett.} \bibinfo{volume}{B718}
  (\bibinfo{year}{2013}) \bibinfo{pages}{795}.
\bibitem[{Abelev et~al.(2013)}]{Abelev:2012ola}
\bibinfo{author}{B.~Abelev}, et~al. (\bibinfo{collaboration}{ALICE
  Collaboration}), \bibinfo{journal}{Phys. Lett.} \bibinfo{volume}{B719}
  (\bibinfo{year}{2013}) \bibinfo{pages}{29}.
\bibitem[{Aad et~al.(2013)}]{Aad:2013fja}
\bibinfo{author}{G.~Aad}, et~al. (\bibinfo{collaboration}{ATLAS
  Collaboration}), \bibinfo{journal}{Phys. Lett.} \bibinfo{volume}{B725}
  (\bibinfo{year}{2013}) \bibinfo{pages}{60}.
\bibitem[{Adare et~al.(2013)}]{Adare:2013piz}
\bibinfo{author}{A.~Adare}, et~al. (\bibinfo{collaboration}{PHENIX
  Collaboration}), \bibinfo{journal}{Phys. Rev. Lett.} \bibinfo{volume}{111}
  (\bibinfo{year}{2013}) \bibinfo{pages}{212301}.
\bibitem[{Bo\.zek(2012)}]{Bozek:2011if}
\bibinfo{author}{P.~Bo\.zek}, \bibinfo{journal}{Phys. Rev.}
  \bibinfo{volume}{C85} (\bibinfo{year}{2012}) \bibinfo{pages}{014911}.
\bibitem[{Bo\.zek and Broniowski(2013)}]{Bozek:2013uha}
\bibinfo{author}{P.~Bo\.zek}, \bibinfo{author}{W.~Broniowski},
  \bibinfo{journal}{Phys. Rev.} \bibinfo{volume}{C88} (\bibinfo{year}{2013})
  \bibinfo{pages}{014903}.
\bibitem[{Bzdak et~al.(2013)Bzdak, Schenke, Tribedy, and
  Venugopalan}]{Bzdak:2013zma}
\bibinfo{author}{A.~Bzdak}, \bibinfo{author}{B.~Schenke},
  \bibinfo{author}{P.~Tribedy}, \bibinfo{author}{R.~Venugopalan},
  \bibinfo{journal}{Phys. Rev.} \bibinfo{volume}{C87} (\bibinfo{year}{2013})
  \bibinfo{pages}{064906}.
\bibitem[{Qin and M{\"u}ller(2014)}]{Qin:2013bha}
\bibinfo{author}{G.-Y. Qin}, \bibinfo{author}{B.~M{\"u}ller},
  \bibinfo{journal}{Phys. Rev.} \bibinfo{volume}{C89} (\bibinfo{year}{2014})
  \bibinfo{pages}{044902}.
\bibitem[{Kozlov et~al.(2014)Kozlov, Luzum, Denicol, Jeon, and
  Gale}]{Kozlov:2014fqa}
\bibinfo{author}{I.~Kozlov}, \bibinfo{author}{M.~Luzum},
  \bibinfo{author}{G.~Denicol}, \bibinfo{author}{S.~Jeon},
  \bibinfo{author}{C.~Gale}  (\bibinfo{year}{2014}).
  \href{http://arxiv.org/abs/1405.3976}{arXiv:1405.3976}.
\bibitem[{Werner et~al.(2014)Werner, Bleicher, Guiot, Karpenko, and
  Pierog}]{Werner:2013ipa}
\bibinfo{author}{K.~Werner}, \bibinfo{author}{M.~Bleicher},
  \bibinfo{author}{B.~Guiot}, \bibinfo{author}{I.~Karpenko},
  \bibinfo{author}{T.~Pierog}, \bibinfo{journal}{Phys. Rev. Lett.}
  \bibinfo{volume}{112} (\bibinfo{year}{2014}) \bibinfo{pages}{232301}.
\bibitem[{Nagle et~al.(2014)Nagle, Adare, Beckman, Koblesky, Koop
  et~al.}]{Nagle:2013lja}
\bibinfo{author}{J.~Nagle}, \bibinfo{author}{A.~Adare},
  \bibinfo{author}{S.~Beckman}, \bibinfo{author}{T.~Koblesky},
  \bibinfo{author}{J.~O. Koop}, et~al., \bibinfo{journal}{Phys.Rev.Lett.}
  \bibinfo{volume}{113} (\bibinfo{year}{2014}) \bibinfo{pages}{112301}.
\bibitem[{He et~al.(2015)He, Edmonds, Lin, Liu, Molnar et~al.}]{He:2015hfa}
\bibinfo{author}{L.~He}, \bibinfo{author}{T.~Edmonds}, \bibinfo{author}{Z.-W.
  Lin}, \bibinfo{author}{F.~Liu}, \bibinfo{author}{D.~Molnar}, et~al.
  (\bibinfo{year}{2015}).
  \href{http://arxiv.org/abs/1502.05572}{arXiv:1502.05572}.
\bibitem[{Ma and Bzdak(2014)}]{Ma:2014pva}
\bibinfo{author}{G.-L. Ma}, \bibinfo{author}{A.~Bzdak},
  \bibinfo{journal}{Phys.Lett.} \bibinfo{volume}{B739} (\bibinfo{year}{2014})
  \bibinfo{pages}{209--213}.
\bibitem[{Bzdak and Ma(2014)}]{Bzdak:2014dia}
\bibinfo{author}{A.~Bzdak}, \bibinfo{author}{G.-L. Ma},
  \bibinfo{journal}{Phys.Rev.Lett.} \bibinfo{volume}{113}
  (\bibinfo{year}{2014}) \bibinfo{pages}{252301}.
\bibitem[{Koop et~al.(2015)Koop, Adare, McGlinchey, and Nagle}]{Koop:2015wea}
\bibinfo{author}{J.~D.~O. Koop}, \bibinfo{author}{A.~Adare},
  \bibinfo{author}{D.~McGlinchey}, \bibinfo{author}{J.~Nagle}
  (\bibinfo{year}{2015}).
  \href{http://arxiv.org/abs/1501.06880}{arXiv:1501.06880}.
\bibitem[{Dumitru and Giannini(2014)}]{Dumitru:2014dra}
\bibinfo{author}{A.~Dumitru}, \bibinfo{author}{A.~V. Giannini},
  \bibinfo{journal}{Nucl.Phys.} \bibinfo{volume}{A933} (\bibinfo{year}{2014})
  \bibinfo{pages}{212--228}.
\bibitem[{Dumitru et~al.(2014)Dumitru, McLerran, and Skokov}]{Dumitru:2014yza}
\bibinfo{author}{A.~Dumitru}, \bibinfo{author}{L.~McLerran},
  \bibinfo{author}{V.~Skokov}  (\bibinfo{year}{2014}).
  \href{http://arxiv.org/abs/1410.4844}{arXiv:1410.4844}.
\bibitem[{Kovchegov and Wertepny(2013)}]{Kovchegov:2012nd}
\bibinfo{author}{Y.~V. Kovchegov}, \bibinfo{author}{D.~E. Wertepny},
  \bibinfo{journal}{Nucl.Phys.} \bibinfo{volume}{A906} (\bibinfo{year}{2013})
  \bibinfo{pages}{50--83}.
\bibitem[{Dusling and Venugopalan(2013)}]{Dusling:2013oia}
\bibinfo{author}{K.~Dusling}, \bibinfo{author}{R.~Venugopalan},
  \bibinfo{journal}{Phys. Rev.} \bibinfo{volume}{D87} (\bibinfo{year}{2013})
  \bibinfo{pages}{094034}.
\bibitem[{Bo\.zek et~al.(2014)Bo\.zek, Bzdak, and Skokov}]{Bozek:2013sda}
\bibinfo{author}{P.~Bo\.zek}, \bibinfo{author}{A.~Bzdak},
  \bibinfo{author}{V.~Skokov}, \bibinfo{journal}{Phys.Lett.}
  \bibinfo{volume}{B728} (\bibinfo{year}{2014}) \bibinfo{pages}{662}.
\bibitem[{Gelis et~al.(2010)Gelis, Iancu, Jalilian-Marian, and
  Venugopalan}]{Gelis:2010nm}
\bibinfo{author}{F.~Gelis}, \bibinfo{author}{E.~Iancu},
  \bibinfo{author}{J.~Jalilian-Marian}, \bibinfo{author}{R.~Venugopalan},
  \bibinfo{journal}{Ann.Rev.Nucl.Part.Sci.} \bibinfo{volume}{60}
  (\bibinfo{year}{2010}) \bibinfo{pages}{463--489}.
\bibitem[{Granier~de Cassagnac(2014)}]{GranierdeCassagnac:2014jha}
\bibinfo{author}{R.~Granier~de Cassagnac} (\bibinfo{collaboration}{CMS
  Collaboration}), \bibinfo{journal}{Nucl.Phys.} \bibinfo{volume}{A931}
  (\bibinfo{year}{2014}) \bibinfo{pages}{13--21}.
\bibitem[{Chojnacki et~al.(2012)Chojnacki, Kisiel, Florkowski, and
  Broniowski}]{Chojnacki:2011hb}
\bibinfo{author}{M.~Chojnacki}, \bibinfo{author}{A.~Kisiel},
  \bibinfo{author}{W.~Florkowski}, \bibinfo{author}{W.~Broniowski},
  \bibinfo{journal}{Comput. Phys. Commun.} \bibinfo{volume}{183}
  (\bibinfo{year}{2012}) \bibinfo{pages}{746}.
\bibitem[{Bo\.zek(2010)}]{Bozek:2009dw}
\bibinfo{author}{P.~Bo\.zek}, \bibinfo{journal}{Phys. Rev.}
  \bibinfo{volume}{C81} (\bibinfo{year}{2010}) \bibinfo{pages}{034909}.
\bibitem[{Bo\.zek et~al.(2013)Bo\.zek, Broniowski, and
  Torrieri}]{Bozek:2013ska}
\bibinfo{author}{P.~Bo\.zek}, \bibinfo{author}{W.~Broniowski},
  \bibinfo{author}{G.~Torrieri}, \bibinfo{journal}{Phys. Rev. Lett.}
  \bibinfo{volume}{111} (\bibinfo{year}{2013}) \bibinfo{pages}{172303}.
\bibitem[{Adam et~al.(2015)}]{Adam:2015pya}
\bibinfo{author}{J.~Adam}, et~al. (\bibinfo{collaboration}{ALICE
  Collaboration})  (\bibinfo{year}{2015}).
  \href{http://arxiv.org/abs/1502.00559}{arXiv:1502.00559}.
\bibitem[{Teaney(2003)}]{Teaney:2003kp}
\bibinfo{author}{D.~Teaney}, \bibinfo{journal}{Phys. Rev.}
  \bibinfo{volume}{C68} (\bibinfo{year}{2003}) \bibinfo{pages}{034913}.
\bibitem[{Wang and Gyulassy(1991)}]{Wang:1991hta}
\bibinfo{author}{X.-N. Wang}, \bibinfo{author}{M.~Gyulassy},
  \bibinfo{journal}{Phys.Rev.} \bibinfo{volume}{D44} (\bibinfo{year}{1991})
  \bibinfo{pages}{3501--3516}.
\bibitem[{Li and Ko(1995)}]{Li:1995pra}
\bibinfo{author}{B.-A. Li}, \bibinfo{author}{C.~M. Ko},
  \bibinfo{journal}{Phys.Rev.} \bibinfo{volume}{C52} (\bibinfo{year}{1995})
  \bibinfo{pages}{2037--2063}.

\end{thebibliography}

\end{document}